\documentclass[12pt]{article}
\title { \bf Bubble formation in $\phi^6$ potential}
\author{Hatem Widyan\thanks{E--mail : hwidyan@ictp.it, widyan@ahu.edu.jo }\\
    {Department of Physics} \\
    {Al-Hussein Bin Talal University, Ma'an, Jordan}
    }

\begin {document}
\maketitle
%
%
%
{\bf Abstract:} Scalar field theory with an asymmetric potential
is studied at zero temperature and high-temperature for $\phi^6$
potential. The equations of motion are solved numerically to
obtain O(4) spherical symmetric and O(3) cylindrical symmetric
bounce solutions. These solutions control the rates for tunneling
from the false vacuum to the true vacuum by bubble formation. The
range of validity of the thin-wall approximation (TWA) is
investigated. An analytical solution for the bounce is presented,
which reproduces the action in the thin-wall as well as the
thick-wall limits.


{\bf keywords:} phase transition, tunneling, scalar field theory
%
%
\begin {section}{\bf Introduction}
\par
The problem of decay of a metastable state via quantum tunneling
has important
applications in many branches of physics, from condensed matter
to particle
physics and cosmology. The tunneling is not a perturbative effect.
In the semi-classical approximation, the decay rate per unit volume is
given by an expression of the form
\begin{equation}
  \Gamma \quad = \quad A \quad e^{-S_E} , \label{eq:decay}
\end{equation}
where $S_E$ is the Euclidean action for the bounce:
the classical solution of the equation of motion with appropriate boundary
conditions. The bounce has turning points at the configurations at which
the system enters and exits the potential barrier, and analytic
continuation to Lorentzian time at the exit point gives us the
configuration of the system at that point and its subsequent evolution.
The solution of the
equation of motion looks like a bubble in four dimensional Euclidean space
with radius R and thickness proportional to the coefficient of the
symmetry breaking term in the potential. When there are more than
one solution
satisfying the boundary conditions, the one with the lowest $S_E$
dominates equation (\ref{eq:decay}).
 The prefactor A comes from Gaussian functional integration over small
fluctuations around the bounce. The zero-temperature formalism is
well-developed \cite{Langer,Coleman,Glasser}. In particular, it has been
proved rigorously that the least action is given by the bounce which
is O(4) invariant \cite{Glasser}.

 Linde \cite{Linde} extended the formalism to finite temperatures. He
suggested that at temperatures much smaller than the inverse radius of the
bubble at zero-temperature, the bounces are periodic in the Euclidean
time $\tau$
direction and widely separated. Beyond this temperature they start
merging into one another producing what is known as
``wiggly cylinder'' solutions. As one keeps
increasing the temperature these wiggles smoothly straighten out, and the
solution goes into an O(3) invariant cylinder (independent of
Euclidean time $\tau$) solution that dominates the thermal
activation regime.

 A numerical and analytical calculations of the first and second
 order phase transitions has been considered by many authors. For
 example, an analytical calculation of the nucleation rate for
 first order phase transitions beyond the TWA for the standard
 Ginzburg-Landau potential with $\phi$ asymmetric term
 has been studied by M$\ddot{\rm u}$nster and Rotsch \cite{munster}.
 We have considered in an earlier work the $\phi^4$ theory with
different  symmetry breaking terms \cite{Hatem}, where we have
obtained numerical as well as analytical solution for different
values of the asymmetric term. In this paper we consider $\phi^6$
potential motivated by the recent work on baryon asymmetry in the
standard model with a low cut-off \cite{bodeker}. Also, if the
Higgs potential is stabilized by a $\phi^6$ interaction, a strong
first order transition can occur for Higgs masses well above $100$
Gev \cite{zhang,ham,grojean}. Moreover, the $\phi^6$ potential has
been investigated by many authors in the context of condensed
matter as well as particle physics (see for example
\cite{Bergner,Amaral,Flores,Joy,Arnold,Zamo,Lu,Kim}).

The general form of the potential is

\begin{equation}
 U(\phi) = {1 \over 2} m^2 \phi^2 +\lambda  \phi^4 + g
\phi^6 , \label{bergner1}
\end{equation}
which has a second-order transition in $m^2$ if $\lambda >0$ by
ignoring corrections due to fluctuations, a first-order transition
in $m^2$ if $\lambda < 0$, and a tricritical point at $\lambda=0$
\cite{Arnold}. Since we are interested in the case of getting
bounce solution, so we take the case of $\lambda < 0$. Following
\cite{Bergner}, we rewrite the potential $U(\phi)$ in terms of the
parameters $\phi_0$ and $\delta$ such that
\begin{equation}
 U(\phi) = g \, \phi^2 (\phi^2-\phi_0^2)^2-\delta \phi^2,
 \label{bergner2}
\end{equation}
where $\phi_0^2=-\lambda/{2g}$ and
$\delta=(\lambda^2/g-2m^2)/4$.
  Looking carefully to the potential, we realize that by fixing
$\lambda$ and $g$, then $\delta$ is changed by changing the value
of $m^2$. Hence the $\delta$ term plays the asymmetric part of the
potential and is responsible for the first-order feature of the
phase transition, by causing the coexistence of two minima (false
and true) separated by a barrier. So, for different values of
$\delta$, we get different shape of the potential.

An interesting special case is the so called thin-wall
approximation (TWA), when the bubble radius $R$ is much larger
than the thickness of the bubble wall and the barrier between the
two minima is large. In this limit ($\delta \rightarrow 0$), there
is an analytical formula for $S_E$ in terms of the wall surface
energy, and the details of the field theory are unimportant.
However, it would be nice to also have an analytical interpolating
form for the solution itself.  Also, it is not clear a priori what
the limit of validity of the TWA is. Another interesting case is
called the thick wall which is reached when the barrier is small. One can
easily show that the barrier is completely disappeared when
$\delta=g\phi_0^4$ and in this case there is no bubbles formed and the
field goes from the false vacuum to the true vacuum without tunneling.

  In this paper we address the above issues. We obtain accurate
numerical solutions for the zero-temperature and high-temperature
bounces for $\phi^6$ theory with $\phi^2$ symmetry-breaking term.
We compute the actions in each case, and find that, for a modest
value of the asymmetric coupling $\delta(=0.1)$, the action given
by TWA formula agrees to within $12.8 \%$ with that obtained from
the numerical solution. We test the criterion for the
goodness of TWA, in terms of the temperature $T_\beta$ at which
the actions of the O(4) and O(3) solutions become equal
\cite{Hatem}. A numerical investigation shows that the TWA holds up to
$\delta \sim 0.25$. Finally, we present an analytical solution which
satisfies the equation of motion with parameters fixed by
demanding stationary action. This reproduces TWA results very well
and, in the thick-wall limit, is in good agreement with the
numerical results.

\end{section}
\begin {section}{ \bf Bubble formation}
%
%
\indent Let us consider a scalar field theory with a Lagrangian density
\begin{equation}
\mathcal{L}(\phi)  =  {1 \over 2}({\partial_{\mu}\phi})^2 -
U(\phi) ,
\end{equation}
 where the potential $U(\phi)$ has two minima at
$\phi_-$ (false vacuum) and $\phi_+$ (true vacuum).

 In the semi-classical approximation the barrier tunneling
leads to the appearance of
bubbles of a new phase with $\phi = \phi_+$ as classical solutions
in Euclidean space (i.e., imaginary time $\tau$). To calculate the
probability
of such a process in quantum field theory at zero temperature, one should
first solve the Euclidean equation of motion :
\begin{equation}
 \partial_{\mu}\partial_{\mu}\phi = {dU(\phi) \over d\phi}
\label{equa:motion} ,
\end{equation}
with the boundary condition
$\phi \to \phi_-$ as $ \vec x^2+\tau^2 \to \infty$ ,
where $\tau$ is the imaginary time. The probability of tunneling per unit
time per unit volume is given by
\begin{equation}
\Gamma  =   A \quad  e{^{-S_E[\phi]}} \label{eq:S4},
\end{equation}
where $ S_E[\phi]$ is the Euclidean action corresponding to
the solution of equation (\ref{equa:motion}) and given by the
following expression :
\begin{equation}
S_E[\phi] =  \int d^4{x} \left[ {1 \over 2} ({\partial\phi \over
\partial\tau})^2 + {1 \over 2} ( \nabla\phi)^2 + U(\phi) \right]
. \label{sephi}
\end{equation}
\indent It is sufficient to restrict ourselves to the O(4) symmetric
solution $ \phi(\vec x^2+\tau^2)$ , since it is this solution that
provides the minimum of the action $ S_E[\phi]$ \cite{Glasser}.
 In this case equation (\ref{equa:motion}) takes the simpler form
\begin{equation}
{d^2\phi \over d\rho^2} + {3 \over \rho} {d\phi \over d\rho}
= {dU(\phi) \over d\phi } \label{equt:S4},
\end{equation}
where $\rho=\sqrt{\vec x^2+\tau^2}$, with boundary conditions
\begin{equation}
\phi \to \phi_-  \quad {\rm as} \quad \rho \to \infty ,\quad
{d\phi \over d\rho }= 0 \quad {\rm at} \quad \rho = 0 \label{cond:1}.
\end{equation}
 We denote the action of this solution by $S_4$.

\indent Now let us consider the finite temperature case. Following
\cite{Linde}, in the calculation of the action $S_E(\phi)$ the
integration over $\tau$ is reduced simply to multiplication by
$T^{-1}$, i.e., $S_E[\phi]= T^{-1} \> S_3[\phi]$. Here
$S_E[\phi]$ is the four-dimensional action and $S_3[\phi]$
is the three-dimensional action corresponding to the
O(3)-symmetric bubble and given by :
\begin{equation}
 S_3[\phi]  =   \int d^3 r \left[ {1 \over 2} {(\nabla \phi)^2 }
+ U(\phi,T) \right]  \label{num61:61} .
\end{equation}
\indent To calculate $S_3(\phi)$ it is necessary to solve the equation
\begin{equation}
{d^2\phi \over dr^2} + {2 \over r }{d\phi \over dr}=
{dU(\phi,T) \over d\phi } \label{equ:S3}
\end{equation}
with boundary conditions
\begin{equation}
\phi \to \phi_- \quad {\rm as} \quad r \to \infty , \quad
{d\phi \over dr} = 0 \quad {\rm at} \quad r=0.
\end{equation}
where $r = \sqrt{\vec x^2}$. The complete expression for the
probability of tunneling per unit time per unit volume in the
high-temperature limit ($T >> R^{-1}$) is obtained in analogy to
the one used in \cite{Coleman} and is given by:
\begin{equation}
\Gamma(T)  =  A(T) \> e^{-S_3[\phi,T]/T} .
\end{equation}
 In the theory of bubble formation , the interesting quantity
to calculate is the probability of decay between $\phi=\phi_-$
and $\phi= \phi_+$
which are the two minima of $U(\phi)$. There is an interesting case
 (in the sense that the action can be calculated analytically) when
$U(\phi_+)-U(\phi_-)= \varepsilon$ is much smaller than
the height of the barrier. This is known as the thin-wall approximation
(TWA). At $T=0$, in the TWA limit, the action $S_4$ of the
O(4)-symmetric bubble is equal to
\begin{equation}
S_4={{27 \pi^2 S_1^4} \over {2 \varepsilon^3}} . \label{ac:ac}
\end{equation}

 Here $S_1$ is the bubble wall surface energy (surface tension), given by
\begin{equation}
 S_1 = \int_0^\infty d\rho \left[({d\phi \over d\rho})^2 +
U(\phi)\right] , \label{a:S_1}
\end{equation}
and the integral should be calculated in the limit $\varepsilon \to
0$. The bubble radius $R$ is written in terms of $S_1$ and
$\varepsilon$ as
\begin{equation}
R= {{3 S_1} \over \varepsilon}.
\end{equation}
 The results presented above were obtained by Coleman \cite{Coleman}.

 These results can be easily extended to the case $ T >> R^{-1}$
\cite{Linde}. To this end it is sufficient to take into account that
\begin{eqnarray}
S_3 & = & 4 \pi \int_0^\infty dr \> r^2 \left[ {1 \over 2}
({d\phi \over dr})^2 + U(\phi,T) \right]  \nonumber \\
 & = & {-{4 \over 3}} \varepsilon \pi R(T)^3 + 4 \pi R(T)^2 S_1(T)
 \label{equation:S_3} ,
\end{eqnarray}
where $S_1(T)$ is the bubble wall surface energy (surface tension)
at finite temperature and is given by:
\begin{equation}
S_1(T) = \int_0^\infty dr \left[ ({d\phi \over dr})^2 + U(\phi,T)
\right].
\end{equation}
 As before, the integral should be calculated in the limit
$\varepsilon \to 0$.

The bubble radius $R(T)$ is calculated by minimizing
$S_3$ with respect to $R(T)$ and this gives us
\begin{equation}
R(T)= {{2 S_1(T)} \over \varepsilon} ,
\end{equation}
whence it follows that
\begin{equation}
S_3={{16 \pi S_1^3(T)} \over {3 \varepsilon^2}} . \label{ac1:ac1}
\end{equation}

\end{section}
%
%
%
\begin {section}{\bf Numerical results}
For O(4) symmetry at $T=0$, equation (\ref{sephi}) reduces to
\begin{equation}
S_4= 2 \pi^2 \int_0^\infty d\rho \> \rho^3 \left[ {1 \over 2 }
({d\phi \over d\rho })^2 + U(\phi) \right] . \label{num2:2}
\end{equation}
 We compute the action for different values of the parameter $\delta$ in the
symmetry-breaking term in the potential $U(\phi)$,
equation (\ref{bergner2}), which reads as

\begin{equation}
S_4= 2 \pi^2  \int_0^\infty d\rho \> \rho^3 \left[ {1 \over 2} ({ d\phi
\over d\rho})^2 +g \, \phi^2 (\phi^2-\phi_0^2)^2-\delta
\phi^2 \right] . \label{num5:5}
\end{equation}
Following \cite{Bergner}, we assume $\phi_0=2.39$ and $g=0.07$, then
the only adjustable parameter in the Lagrangian is $\delta$. So, by
covering the whole range $0 < \delta < g\phi_0^4$ we should be
covering all relevant cases.

The equation of motion is now
\begin{equation}
{d^2\phi \over d\rho^2}+ {3 \over \rho} {d\phi \over d\rho}
= 6g \phi^5-8g \phi_0^2 \phi^3+ 2( g \phi_0^4-\delta)
\phi , \label{num6:6}
\end{equation}
and the boundary conditions are
\begin{equation}
\phi=0 \quad {\rm as} \quad \rho \to \infty ,\quad {d\phi
\over d\rho}=0 \quad {\rm at} \quad \rho=0 .
\end{equation}
By solving equation (\ref{num6:6}) numerically for different values of
$\delta$, substituting the solution in equation (\ref{num5:5})
 and integrating, we obtain the action for each value of
$\delta$.

 At high temperature, we look for the O(3) symmetric solution with
cylindrical symmetry. Then equation (\ref{num61:61}) takes the form
\begin{equation}
S_3=4 \pi \int_o^\infty dr \> r^2 \left[{1 \over 2} ({d\phi
\over dr }) ^2 +   g \, \phi^2 (\phi^2-\phi_0^2)^2-\delta
\phi^2 \right] . \label{num8:8}
\end{equation}
The equation of motion is then
\begin{equation}
{ d^2\phi \over dr^2 } +{2 \over r} { d\phi \over dr } = 6g
\phi^5-8g \phi_0^2 \phi^3+ 2( g \phi_0^4-\delta)
\phi , \label{num9:9}
\end{equation}
and the boundary conditions are
\begin{equation}
\phi=0 \quad {\rm as} \quad r \to \infty ,\quad {d\phi \over
dr }=0 \quad {\rm at} \quad r=0 . \label{bcs3}
\end{equation}
 Again, we solve equation (\ref{num9:9}) numerically for different
values of $\delta$, substitute the solution in equation
(\ref{num8:8}) and integrate to obtain the action for each
$\delta$. Figure 1 shows the bubble profile for different values
of $\delta$. Note that the value of the scalar field $\phi$ inside
the bubble decreases with $\delta$. In figure 2 we have plotted
this value together with the minimum of the potential $U(\phi)$.
At $\delta=0$ the value of $\phi_0$, i.e. $\phi(r=0)$, coincides
with the minimum of the potential $\phi_m$. However, as $\delta$
decreases, the minimum increases while $\phi_0$ initially
increased then it decreases and moves away from the minimum of
$U(\phi)$. Same behavior has been obtained and explained by
\cite{Megevand} and it is due the decreasing of the height of the
potential and the increasing in the energy difference between
minima. So, physically this means that as the barrier between
minima disappears, it becomes easier to from a large bubble with a
small value of $\phi$ inside it. Same result can been obtained
also for the case of zero temperature.

\begin{center}
{\bf Table 1.} Numerical values of the action $T=0$ and
high temperature for different values of the asymmetry parameter
$\delta$.
\end{center}
\begin{center}
\begin{tabular}
{p{1.2cm} p{3.2cm} p{3.2cm}  p{3.2cm}  p{3.2cm}|}
\hline
 $\delta$ & $ S_4$ (Numerical) & $  S_3$ (Numerical)  \\
\hline
 0.1 & 70978.1 & 1620.08  \\
0.2 & 9739.27 & 441.89   \\
0.29 & 3625.65 & 225.41  \\
0.4 & 1519.96 & 127.73   \\
0.6 & 523.12 & 63.36   \\
0.8 & 253.85 & 38.72   \\
1.0 & 143.42 & 26.27   \\
1.2 & 91.07 & 18.98   \\
1.4 & 60.85 & 14.08   \\
1.6 & 42.67 & 10.43   \\
 1.8 & 30.03 & 7.79   \\
2.0 & 21.34 & 5.53   \\
2.2 & 15.01 & 3.52   \\
2.28 & 12.84 & 2.89   \\
\hline
\end{tabular}
\end{center}
\vskip 0.4cm

 As we discussed in the introduction, for small values of $\delta$ we can use the
TWA formula for computing the action. From equation (\ref{a:S_1})
\begin{eqnarray}
S_1 & = & \int_0^\infty dr \left[ ({d\phi \over dr})^2 + g \,
\phi^2 (\phi^2-\phi_0^2)^2 \right] \nonumber \\
& = & - \int_{0}^{\phi_0} d\phi \> \sqrt{2 g \, \phi^2 (\phi^2-\phi_0^2)^2} \nonumber \\
 & = & {3.05}  \label{10:10},
\end{eqnarray}
for $\phi_0=2.39$ and $g=0.07$, see \cite{Bergner}.
 The radius is given by
\begin{equation}
R = {{3 S_1} \over \varepsilon },
\end{equation}
where $ \varepsilon = \phi_0^2 \, \delta$ (see
\cite{Coleman,Bergner}). For $\delta=0.1$, we have $ R=16.02$ and the
value of the action is (see equation (\ref{ac:ac}))
\begin{equation}
S_4 = 61918.4 .  \label{num11:11}
\end{equation}
Comparing this analytical value with the numerical one for
$\delta=0.1$, we get an error equal $12.8\%$. In \cite{Bergner}, the
authors choose $\delta=0.29$ to represents the TWA and they have
concluded that it is not a good value to be taken. This is also
confirmed by our calculations where we get an error approximately
$30\%$.

At high temperature, again $S_1= 3.05$. The value of $R(T)= 10.68$
at $\delta=0.1$ and the action is (see equation (\ref{ac1:ac1}))
\begin{equation}
S_3(T)= 1454.26 . \label{num12:12}
\end{equation}
 Comparing this analytical value with the numerical one for
$\delta=0.1$, we get an error equal $10.24 \%$. Thus even for
$\delta$ as small as $0.1$ the TWA formula for the action does not
give very accurate results. Obviously, there is no point in
comparing numerical results obtained for higher values of $\delta$
with the TWA formula.

  To test our numerical method (we have used Hamming's modified
predictor-corrector method for solving the equation of motion), we
have calculated the action $S_4$ for small values of the symmetry
breaking parameter $\delta$ in the potential and compared it with
the TWA formula. In figure 3, we plot the percentage error in the
TWA formula as a function of $\delta$. The crosses represent our
results while the solid line shows a fit to the data. We see that
the error decreases for small $\delta$, as expected, and
approaches zero as $\delta \rightarrow 0$.

 As already mentioned, at zero temperature the O(4) symmetric solution
has the lowest value of $S_E$, i.e., $S_E=S_4$. At high temperature,
we have $S_E=S_3/T$. At intermediate temperatures other solutions
exist. In the TWA, however, it has been shown \cite{Garriga} that all
other solutions have higher Euclidean action. This corresponds to a
first order phase transition from quantum tunneling at low temperature
to thermal hopping at high temperatures. The transition temperature
$T _\beta$ is given by equating $S_4$ with $S_3/T$, i.e.,
\begin{equation}
T_\beta = {S_3 \over S_4} \label{num15:15}
\end{equation}
If the surface tension $S_1$ is temperature independent, we have
\begin{equation}
S_4  =  {{27 \pi^2 S_1^4} \over {2 \varepsilon^3}} \label{num16:16}
\end{equation}
\begin{equation}
S_3 =  {{16 \pi S_1^3} \over {3 \varepsilon^2}} \label{num17:17}
\end{equation}
Dividing  equation (\ref{num16:16}) by equation (\ref{num17:17}) and putting
$\varepsilon=\phi_0^2  \, \delta$ (see \cite{Coleman}) we get
\begin{equation}
T_\beta = C * \delta
\end{equation}
where
\begin{equation}
C = {32 \, \phi_0^2 \over {81 \pi S_1}}
\end{equation}
Thus we see that, in the TWA, $T_\beta$ increases linearly with
$\delta$. We test this by computing $S_3/S_4$ from our numerical
solutions at different values of $\delta$. Figure 4 shows our
results for the potential given by equation (\ref{bergner2}). We see
that, for $\delta \leq 0.2$, there is very good agreement with the
predicted linear dependence. This also confirms that, in the
domain of validity of the TWA, the surface tension $S_1(T)$ is
independent of $T$. Beyond $\delta \sim 0.2$ in our dimensionless
units, there is a systematic deviation from linearity. Thus we can
say that, for values of $\delta$ larger than this, the wall
thickness becomes important. Same behavior has been obtained also in
our earlier work \cite{Hatem}.

\end{section}
%
\begin{section}{ \bf Analytic solution for zero temperature}
%
%
\def\t{e^{{{(\rho^2 -R^2)}/  \Lambda^2}} + 1}
\def\g{\gamma}
\def\d{\delta}
\def\l{\Lambda}
\def\v0{\phi_0}
 We calculate the action analytically in two extreme limits: the thin-wall and
thick-wall using the potential given by equation (\ref{bergner2}).

\noindent \underline{\it{ Thin-wall limit : $\d \to 0$}}

In an earlier paper \cite{Hatem}, we have found that an analytic
solution for the bounce of the form of a Fermi function:
\begin{equation}
\phi(\rho) =  {\g \over \t} \label,
\end{equation}
is a good approximation for the $\phi^4$ theory. But it has been
shown that for the $\phi^6$ potential, the analytic solution for
the bounce has the form \cite{Flores,Joy}
\begin{equation}
\phi_{\rm wall}^2(\rho)=\frac{\phi_0^2}{1+e^{\mu \rho}},
\end{equation}
where $\mu=\sqrt{8 g}\phi_0^2=4.21$, and $\mu^2$ is the second
derivative of the potential in the TWA limit evaluated at
$\phi_0$. So, motivated by the above results, we assume
\begin{equation}
\phi^2(\rho) =  {\g \over \t} \label{twa1:1},
\end{equation}
where  $\rho=\sqrt{\vec x^2+\tau^2}$, $R$ is the radius of the
bubble and $\l$ its width, acts like a bounce in the TWA and leads
to the correct value for the action $S_4$. The parameter $\sqrt\g$
is approximately equal to true minimum in the TWA. The bounce has
values $\phi = \sqrt\g$ at $\rho=0$ and $0$ at $\rho \to
\infty$. The boundary conditions~(\ref{cond:1}) are satisfied by
equation (\ref{twa1:1}).

To evaluate $\g$, $R$, and $\l$, we substitute the ansatz
(\ref{twa1:1}) in equation (\ref{num6:6}) :
\begin{equation}
{d^2\phi \over d\rho^2}+ {3 \over \rho}{d\phi \over d\rho} =
6g \phi^5-8g \phi_0^2 \phi^3+ 2( g \phi_0^4-\delta)
\phi. \label{twa2:2}
\end{equation}
Then the left-hand side (L.H.S.) and the right-hand side (R.H.S.) are
respectively
\begin{eqnarray}
L.H.S.= {({3 \rho^2 / \l^4})\sqrt\g \over (\t)^{5/2}} & + &{{
(-4 \rho^2 / \l^4 + 4 / \l^2) \sqrt\g} \over (\t)^{3/2} }  \nonumber \\
[0.3cm]
 & + & {{( \rho^2 / \l^4 - 4 / \l^2) \sqrt\g} \over (\t)^{1/2} }
\label{twa3:3} .
\end{eqnarray}
\begin{eqnarray}
R.H.S.= {{6 g \g^{5/2}} \over (\t)^{5/2}}
 & - & {{ 8 g \v0^2 \g^{3/2}} \over (\t)^{3/2} } \nonumber \\ [0.3cm]
 & + & {{2(g\v0^4-\d)\sqrt\g} \over (\t)^{1/2}}
\label{twa4:4} .
\end{eqnarray}
 In the TWA, the solution is constant except in a narrow region near the
wall at $\rho=R$. So, we replace in equation (\ref{twa3:3})
\begin{eqnarray}
\mathrm{ {3 \rho^2 \over \l^4} \,\,\, by \,\,\, {{ 3 R^2} \over \l^4} ( 1-
a \l^2 /R^2) \,\,\, in \,\,\, the \,\,\, { 1 \over (\t)^{5/2}} \,\,\,
term } \label{twa5:5} , \\ [0.3cm]
\mathrm{ {4 \over \l^2} -{4 \rho^2 \over \l^4} \,\,\, by \,\,\, {-}{{ 4 R^2}
\over \l^4} (1- b \l^2 / R^2 ) \,\,\, in \,\,\, the \,\,\, { 1 \over
(\t)^{3/2}}  \,\,\, term }, \label{twa6:6} \\ [0.3cm]
\mathrm{ { \rho^2 \over \l^4}- {4 \over \l^2} \,\,\, by \,\,\, {{
R^2} \over \l^4} (1- c  \l^2 / R^2) \,\,\, in \,\,\, the \,\,\, {
1 \over (\t)^{1/2}} \,\,\, term } \label{twa7:7} ,
\end{eqnarray}
where $a$, $b$ and $c$ are parameters to be determined later.

Comparing equation (\ref{twa3:3}) with equation (\ref{twa4:4}) in the
range
$ R^2(1- \l^2 / R^2)=R^2-\l^2 < \rho^2 < R^2 + \l^2 =R^2(1+ \l^2 / R^2)$
where $\rho^2 \simeq R^2$  as $\l^2 / R^2 << 1$ , we have :
\begin{eqnarray}
2 g \g^2 = {{ R^2} \over \l^4} (1- a \l^2 / R^2) , \nonumber
\\ [0.3cm]
2 g \v0^2 \g = {{R^2} \over \l^4}( 1-b \l^2 / R^2) ,
\label{twa10:10} \\ [0.3cm]
2 (g \v0^4 - \d) = {{R^2 \over \l^4} ( 1- c \l^2 / R^2)} ,
\nonumber
\end{eqnarray}
We can now evaluate the zero-temperature action $S_4$ :
\begin{equation}
S_4= 2 \pi^2 \> \int_0^\infty d\rho \> \rho^3 \left[{1 \over 2}
({d\phi \over d\rho})^2 + U(\phi) \right] .
\label{twa12:12}
\end{equation}
Substituting equation (\ref{twa1:1}) in equation (\ref{twa12:12}) and
integrating we get
\begin{eqnarray}
S_4 & = & 2 \pi^2  R^4  \g \Bigg[ {1 \over {8 \l^2}}+{1\over{4 R^2}}
+{\pi^2 \l^2 \over 24 R^2} + (g\v0^4-\d) \left({1 \over 4}
 +  { \pi^2 \l^4  \over 12 R^4} \right)  \nonumber \\
& & - {1 \over 2} g \v0^2 \g \left( 1- {{2 \l^2}\over  R^2} + {\pi^2
\l^4 \over 3 R^4} \right)
\nonumber \\
& & + { g  \over 4} \g^2 \left( 1- {{3 \l^2} \over { R^2}}+ ({\pi^2
\over 3} + 1) {\l^4 \over R^4} \right) \Bigg].  \label{twa13:13}
\end{eqnarray}
We now determine the parameters $a$, $b$, and $c$ by demanding
${dS_4 / dR}={dS_4 / d\l}={dS_4 / d\g}=0$.
 Differentiating equation (\ref{twa13:13}) and using
equation (\ref{twa10:10}), we find that to
leading order in $\l^2/R^2$,
\begin{eqnarray}
3-2a+4b-2c =0, \nonumber \\ [0.3cm]
2+3a-4b =0 ,  \\ [0.3cm] \label{twa14:14}
-3a+4b-c =0, \nonumber
\end{eqnarray}
which leads to $a=-1$, $b=-1/4$ and $c=2$. Using
equation (\ref{twa10:10}), we can rewrite equation (\ref{twa13:13}) as :
\begin{eqnarray}
S_4 &=& { \pi^2 \over g \v0^2} {R^6 \over \l^6}\Bigg[
\left( \frac{1}{4}-\frac{a}{8}+\frac{b}{4}-\frac{c}{8} \right)   +
\left (\frac{7}{16}+\frac{11 a}{32}-\frac{7
b}{16}-\frac{c}{32} \right) \frac{\l^2}{R^2} \nonumber \\
& & + \left (\frac{3}{32}+\frac{\pi^2}{24}-\frac{a}{32}-\frac{\pi^2
a}{24}-\frac{b}{8}+ \frac{\pi^2 b}{12}-\frac{\pi^2 c}{24} \right)
{\l^4 \over R^4} \nonumber \\
& & + \left( \frac{\pi^2}{96}-\frac{a}{32}-\frac{\pi^2 a}{96}+\frac{\pi^2
b}{48}-\frac{\pi^2 c}{96} \right) \frac{\l^6}{R^6}+
O(\frac{\l^8}{R^8}) \Bigg] , \label{twa15:15}
\end{eqnarray}
This gives
\begin{equation}
S_4 = { \pi^2 \over g \v0^2} {R^6 \over \l^6} \left(
0.063 + 0.141 \frac{\l^2}{R^2} - 0.049 \frac{\l^4}{R^4}-0.020
\frac{\l^6}{R^6 } +  O(\frac{\l^8}{R^8})   \right). \label{twa16:16}
\end{equation}
The quantities $\g$, $R$ and $\l$ are determined from
equation (\ref{twa10:10}) using the values of $a$, $b$, and $c$. So we
have
\begin{equation}
2 g \g^2 - 2 g \v0^2 \g \left({{c-a \over {c-b}}}\right) + 2(g
\v0^4- \d) \left({{b-a \over c-b}}\right) = 0 , \label{twa25:25}\\
[0.3cm]
\end{equation}
which gives
\begin{equation}
{R^2 \over \l^2}= {{g \v0^2 c \g  - b(g \v0^4-\d)} \over {g \v0^2
\g - (g \v0^4- \d)}}, \label{twa26:26}
\end{equation}
and
\begin{equation}
\l^2 = {{b-a} \over {2 g \g^2-2g \v0^2 \g}}={{ {c-b} \over {2 g
\v0^2 \g -2(g \v0^4-\d})}} , \label{twa17:17}
\end{equation}
with $\g$ given by equation (\ref{twa25:25}). We have then, for
$\d=0.1$, $\g=5.83$, which implies that $ R^2/\l^2=35.1$, $R=16.3$,
$\l=2.75$ and  $S_4=70997.3$. Comparing these results with the TWA
formulae, we find that the departure of the radius from the TWA is
$R/R_{TWA}=1.02$ while the departure of the action is ${S_4 /
S_{TWA}}=1.14$, which is a fairly good result. On the other hand, there
is no departure of the radius as well as the action from the numerical
values at $\delta=0.1$ which is an excellent result.
 Table 2 shows our numerical as well as the analytical values of the
action and the radius for different values of $\delta$. We have
calculated the numerical value of the radius when the derivative of
the filed is maximum while in \cite{Megevand} the author has
calculated the radius in a different way.
\begin{center}
{\bf Table 2.} Numerical and analytical values of the
action and the radius for different values of $\delta$.
\end{center}
\begin{center}
\begin{tabular}
{p{1.2cm} p{3.2cm} p{3.2cm}  p{3.2cm}  p{3.2cm}}
\hline
 $\delta$ & $S_4$ (Numerical) & $S_4$ (Analytical) & $R$ (Numerical) &
$R$ (Analytical)  \\
\hline
 0.1 & 70978.1 & 70997.3 & 16.3 & 16.3 \\
0.2 & 9739.27 & 10008.3  & 8.3  & 8.28 \\
0.29 & 3625.65 & 3622.26 & 5.8  & 5.79 \\
0.4 & 1519.96 & 1540.44  & 4.2 & 4.26 \\
0.6 & 523.12 & 543.96  & 2.86  & 2.9 \\
0.8 & 253.85 & 266.95  & 2.2  & 2.22 \\
1.0 & 143.42 & 156.01  & 1.82  & 1.81 \\
1.2 & 91.07 & 101.54 & 1.53  &  1.53 \\
1.4 & 60.85 & 70.97 & 1.32  &  1.33 \\
\hline
\end{tabular}
\end{center}
\vskip 0.4cm

Notice that there is an excellent agreement between the radii while
actions are fairly agree till $\delta=1.0$. So, we conclude that our
ansatz gives us far better results than the TWA formula.
In figure 5 we compare our numerical
result with the analytic one for $\d=0.1$. From the figure we see
that the Fermi function agrees very well with our numerical
results

\vskip 0.4 cm

\noindent

\underline{\it{ Thick-wall limit: $\d \to g\phi^4_0 $}}

 The form of the bounce in equation (\ref{twa1:1}) suggests that the thick
wall limit, which would correspond to small values of $R^2/\l^2$, would
be obtained by approximating the Fermi function by the
Maxwell-Boltzmann function, which leads to a Gaussian:
\begin{equation}
 \phi^2(\rho) = \g e^{-\rho^2/\l^2} . \label{twa19:19}
\end{equation}
 The action for this form of bounce is found to be
\begin{equation}
 S_4 = { 2 {\pi^2 \l^4 \g}} \Bigg[ {1 \over {2\l^2}}
+ { g  \over
18} \g^2 - {1 \over 4}g \v0^2 \g + {1 \over 2}(g \v0^4-\d)\Bigg]. \label{twa20:20}
\end{equation}
Equations (\ref{twa10:10}) then reduce to
\begin{equation}
2 g \g^2 = -{a \over \l^2} ~, \quad
2 g \v0^2 \g  = -{ b \over \l^2} ~, \quad
{ 2(g \v0^4-\d)} = - {c \over\l^2} ~.
\end{equation}
Note that in this case $\g \ll 1$, so $\g^2$ is negligible ($a=0$).

The values of $b$ and $c$ are again obtained by demanding
$dS_4/d\l=dS_4/d\g = 0$. The relation between them is
\begin{eqnarray}
2+b-c=0, \nonumber \\ [0.3cm]
2+b-2c=0. \nonumber
\end{eqnarray}
This gives $b=-2$, $c=0$, giving
\begin{equation}
\d = g \v0^4 , \quad g \v0^2 \g={1 \over \l^2} .
\end{equation}
This yields the action
\begin{equation}
 S_4 = {{\pi^2 \v0^2} \over 2 \d}+ O(\frac{\l^2}{R^2})=12.37
\label{dwt1:1}
\end{equation}
for $\d=2.28$. The numerical value is $S_4=12.84$, so the error is $4.6\%$.

 Thus, the form of the bounce given by equation (\ref{twa10:10}) seems valid
over the whole range of $\d$ (from 0 to 2.28), and in the two extreme
limits is amenable to analytic calculations.

\end{section}
%
%
%
\begin{section}{\bf Analytic solution for high temperature}
\def\ht{e^{{{(r^2 -R^2)}/  \Lambda^2}} + 1}
\def\g{\gamma}
\def\d{\delta}
\def\l{\Lambda}
\def\v0{\phi_0}
We discuss now the high-temperature action $S_3$ for the thin wall limit as
well as thick wall.

\noindent \underline{\it{ Thin-wall limit : $\d \to 0$}}

The bounce takes the following from:
\begin{equation}
\phi^2(r)= {\g \over \ht},
\end{equation}
where $r^2=\vec x^2$ and the other parameters $R$ and $\l$ have the
same physical significance in three dimensions. The boundary conditions
given by equation (\ref{bcs3}) are satisfied by the bounce.

We substitute the bounce in the equation of motion (\ref{num9:9}) and
assume the solution is constant except in a narrow region near the wall
$r=R$. The resulting equations enable us to evaluate the action  given
by equation (\ref{num8:8}), and after integrating we get the following:
\begin{eqnarray}
S_3 & = & 4 \pi R^3 \g \Bigg[ {1 \over {8 \l^2}}+{3\over{16 R^2}}
+{\pi^2 \l^2 \over 64 R^2} + ( g\v0^4-\d) \left({1 \over 3}
 +  { \pi^2 \l^4  \over 24 R^4} \right)  \nonumber \\
& & -  g \v0^2 \g \left( {2 \over 3}- {{ \l^2}\over  R^2} + {\pi^2
\l^4 \over 12 R^4} \right)
+ g  \g^2 \left( {1 \over 3}- {{3 \l^2} \over {4 R^2}}+ ({\pi^2
\over 24} + {1 \over 8}) {\l^4 \over R^4} \right) \Bigg].
\end{eqnarray}
In terms of parameters $a$, $b$ and $c$, the action takes the simpler
form
\begin{eqnarray}
S_3 &=& { 4\pi \over 2 g \v0^2} {R^5 \over \l^6}\Bigg[
\left( \frac{1}{4}-\frac{a}{6}+\frac{b}{3}-\frac{c}{6} \right)   +
\left (\frac{9}{8}-\frac{5 a}{24}+\frac{2
b}{3}-\frac{7c}{12} \right) \frac{\l^2}{R^2} \nonumber \\
& & + \left (\frac{7}{8}+\frac{\pi^2}{64}+\frac{5a}{5}-\frac{\pi^2
a}{48}-\frac{7b}{4}+ \frac{\pi^2 b}{24}-\frac{\pi^2 c}{48} \right)
{\l^4 \over R^4} \nonumber \\
& & + \left( \frac{7\pi^2}{128}-\frac{7a}{32}-\frac{7\pi^2 a}{96}+\frac{7\pi^2
b}{128}-\frac{7\pi^2 c}{96} \right) \frac{\l^6}{R^6}+
O(\frac{\l^8}{R^8}) \Bigg] , \label{atws3}
\end{eqnarray}
where the relations between $a$, $b$, and $c$ to leading order in
$\l^2/R^2$ are
\begin{eqnarray}
1-a+2b-c =0, \nonumber \\ [0.3cm]
1+3a-4b =0 ,  \\ [0.3cm]
-3a+4b+c =0, \nonumber
\end{eqnarray}
which leads to $a=-5$, $b=-7/2$ and $c=-1$. Hence the action in
equation (\ref{atws3}) is reduced to
\begin{equation}
S_3 = { 4\pi \over 2 g \v0^2} {R^5 \over \l^6} \left(
0.083 + 0.417  \frac{\l^2}{R^2} + 0.699  \frac{\l^4}{R^4} + 0.914
\frac{\l^6}{R^6}  + O(\frac{\l^8}{R^8}) \right),
\end{equation}
where $\g$, $\l$ and $R$ are obtained from equations (\ref{twa25:25}),
(\ref{twa26:26}) and (\ref{twa17:17}).

We have then, for
$\d=0.1$, $\g=6.04$, which implies that $ R^2/\l^2=22.62$, $R=11.06$,
$\l=2.33$ and  $S_3=1691.95$. Comparing these results with the TWA
formulae, we find that the departure of the radius from the TWA is
$R/R_{TWA}=1.02$ while the departure of the action is ${S_4 /
S_{TW}}=1.16$, which is a fairly good result. On the other hand, there
is a very small departure of the radius as well as the action from the
numerical values at $\delta=0.1$ which is an excellent
result. Similarly as in the case of zero temperature, if you go to
higher values of $\delta$, then there will be a departure from the
numerical results.

In the TWA the radius of the bubble is much greater than its
thickness. So, for $\delta=0.1$, we get $\l=2.75$ which is much
less than $R=16.3$ as is expected. Same result is obtained for the
zero temperature as well as in \cite{Megevand}.

%
%
\vskip 0.4cm

\noindent \underline{\it{ Thick-wall limit : $\d \to g\phi_0^4$}}

At higher temperature the bounce takes the form as the case of zero
temperature,
\begin{equation}
 \phi^2(r) = \g e^{-r^2/\l^2} , \label{twa119:119}
\end{equation}
with the action
\begin{equation}
 S_3 = {  {\pi \l^3 \sqrt{\pi}} \g} \Bigg[ {3 \over {4\l^2}}
+ { 1  \over
{3^{3/2}}} g \g^2 - {4 \over {2^{5/2}}} g \v0^2 \g +
(g \v0^4-\d)\Bigg].
\label{twa200:200}
\end{equation}
Again defining $2 g \v0^2 \g  = - b/ \l^2$, ${ 2(g \v0^4-\d)} = - c /
\l^2$ and neglecting $\g^2$, we find $b$ and $c$ as before by
demanding $dS_3/d\l=dS_3/d\g=0$. The relation between $b$ and $c$ is
given by
\begin{eqnarray}
{3 \over 16} + {b \over 2^{5/2}} - {c \over 8} =0, \nonumber \\ [0.3cm]
{3 \over 16} + {3b \over 2^{7/2}} - {3c \over 8} =0, \nonumber
\end{eqnarray}
which leads to $b=-\sqrt 2$ and $c=-1/2$, giving $\l^2=1/(4
g\v0^4-4\d)$  and $\g=(4 g \v0^4-4\d)/({\sqrt 2} g \v0^2)$.
The action can be simplified to
\begin{equation}
S_3=\pi {\sqrt{2\pi(g \v0^4-\d)} \over {2 g \v0^2}}+
O(\frac{R^2}{\l^2}). \label{s3twaa}
\end{equation}
Note that the value of the action is independent of $\l$ and
depends only on $\delta$ and if $\d= g \v0^4$, then $\g=0$ and the
action $S_3=0$ which is consistent with the result that the hump
of the potential will disappear at this value of $\d$. Another
important result is that $\l$ will diverge in the limit $\d= g
\v0^4$ which has been also obtained by \cite{Megevand}. So, to get
a real value of action we must always have $\d < g \v0^4$.

We have plotted in figure 6 the numerical and analytical bubbles
for $\d=2.0$. Note that in spite of the discrepancy in the value
of $\g$ for the numerical and analytical profiles which is due to
the neglecting the terms of order $R^2/\l^2$ in equation
(\ref{s3twaa}) the departure of the actions is small, i.e.
$S_3({\rm numerical})/S_3({\rm analytical})=1.05.$


\end{section}

\begin{section}{ Conclusions}
%
%
 We have obtained accurate numerical solutions for the
zero-temperature and high-temperature bounces for
$\phi^6$ potential with $\phi^2$ symmetry-breaking for the entire wall
thickness interval $ 0 \le \delta \le g \phi_0^4$. We compute
the actions in each case and find that, for a modest value of the
asymmetric coupling $\delta=0.2$, the action given by the TWA formula
agrees to within $12.8 \%$ with that obtained from the numerical
solution. At high temperatures, the conclusion is qualitatively similar.

 We have checked our numerical method by comparing the action
obtained numerically with the one obtained from the TWA formula.
Very good agreement is obtained as we go to small values of
$\delta$. We also verify that as $\delta$ is reduced the error in the
TWA formula goes to
zero. We check the criterion for the goodness of TWA proposed in
\cite{Hatem}, in terms of the relation between $\delta$ and
the temperature $T_\beta$ at which the actions of the O(4) and O(3)
solutions become equal. A numerical investigation shows that TWA holds
up to $\delta \sim 0.2$. Finally, we present an analytical solution
which satisfies the equation of motion in an approximate sense in two
limiting cases. The
first of these reproduces the leading corrections to the
TWA results very well and it fairly matches the numerical results
of the action up to
$\delta=1.0$. The second is applicable for the opposite case of a very
thick wall. This gives us insights into the nature of the bounce
solutions for various values of $\delta$ going from thin to thick walls.

Some of our results match very well with those obtained in
\cite{Megevand}. For example, we get the same behavior of the the
minimum of the potential $\phi_m$ and the value of
the $\phi$ inside the nucleated bubble, $\phi_0$, see figure
2. Moreover, the divergence of the thick of wall at the vanishing of
the hump of the wall is obtained in \cite{Megevand} numerically while
we get the same behavior analytically.

Much of the work on inflationary models relies on the zero-temperature
potential, so our results could be relevant for inflation
\cite{inflation}. They may also have some bearing on the formation of
topological defects in a first order phase transition where authors
consider zero-temperature potentials, see for example \cite{Digal}.

 So far, we have discussed the action only at zero and high
temperatures. To obtain the bounce solution at intermediate
temperatures, we have to solve a partial differential equation with
periodic boundary conditions in the $\tau$ direction either
numerically or analytically. This work will be  presented in a future
publication.

\end{section}
\begin{section}*{Acknowledgements}
%
The author would like to thank the abdus salam international center for
theoretical physics for the financial support and warm hospitality where
this work has been done.
\end{section}
\bibliography{plain}

\newpage
\begin{section}*{Figure Caption}
{\bf Figure 1.} Shape of the critical bubble at different
$\delta$.

{\bf Figure 2.} The minimum of the potential $\phi_m$ and the value of
the $\phi$ inside the nucleated bubble, $\phi_0$.

{\bf Figure 3.} Error in the TWA formula as a function of $\delta$.
The crosses represent our results while the solid line shows a fit
to the data.

{\bf Figure 4.}  Deviation of $T_\beta$ from the TWA limit. The dashed
line represents the TWA limit while the crosses are our numerical
results.

{\bf Figure 5.} $\phi$ as a function of $\rho$. The dashed line is the
Fermi function while the doted line is the numerical result.

{\bf Figure 6.} $\phi$ as a function of $r$. The dashed line is
the Gaussian function while the solid line is the numerical
result.
\end{section}

\newpage
\begin{figure}[ht]
\vskip 15truecm

\includegraphics{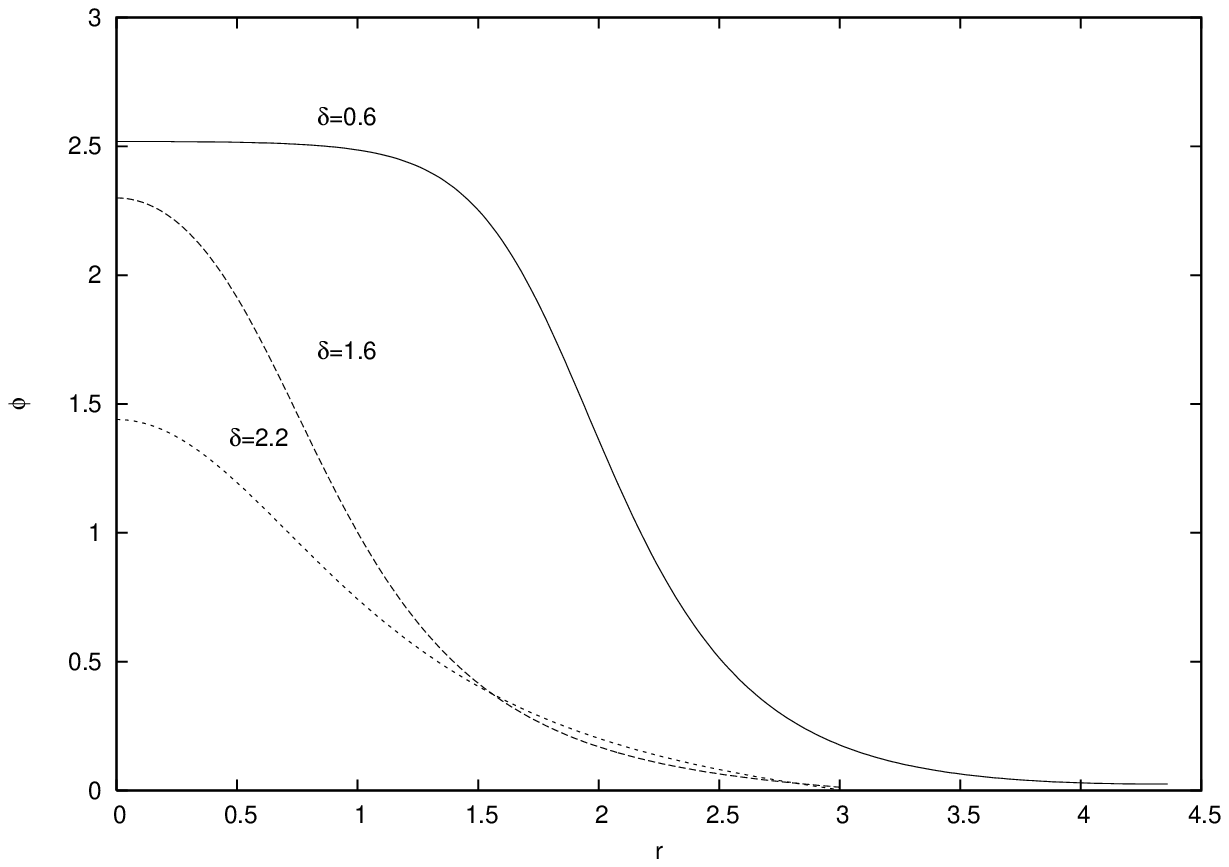}

\caption{}
\end{figure}
\newpage
\begin{figure}[ht]
\vskip 15truecm
 \includegraphics{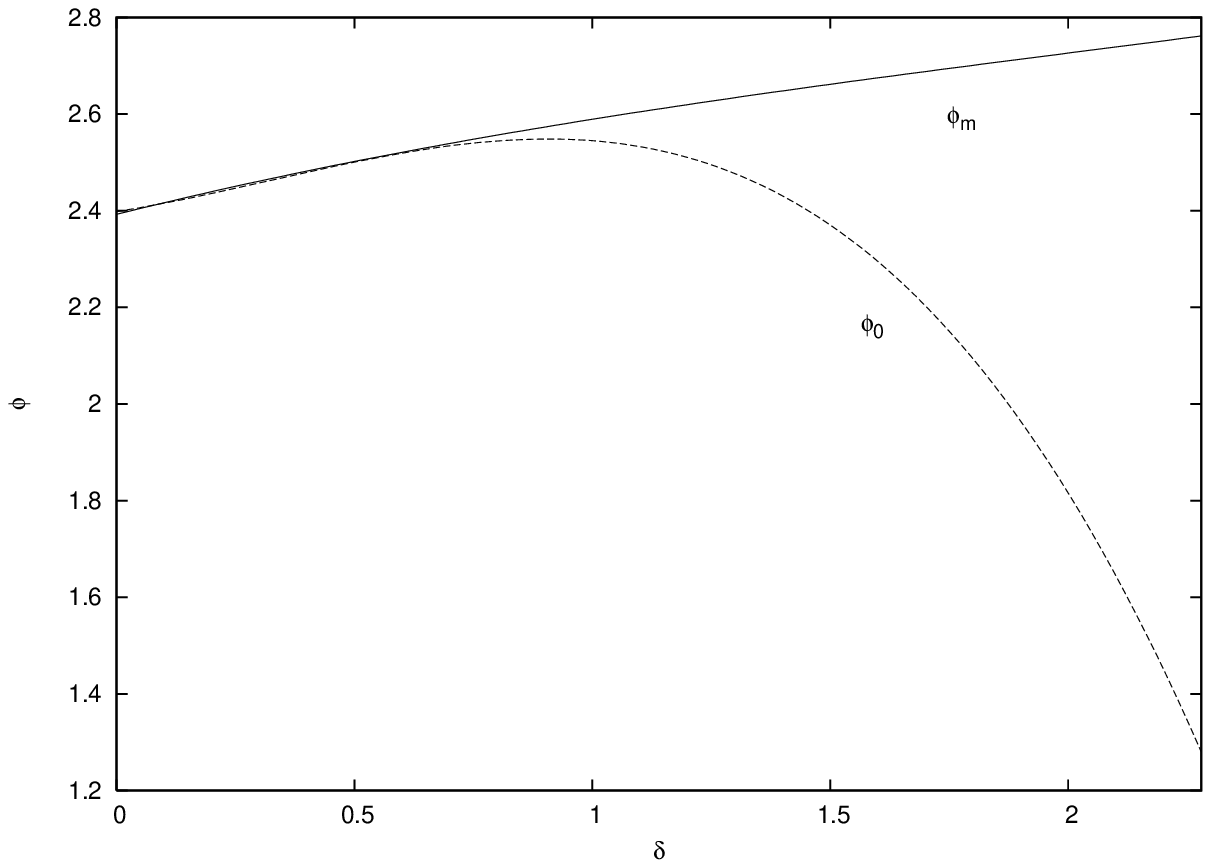}
\caption{}
\end{figure}
\newpage
\begin{figure}[ht]
\vskip 15truecm
 \includegraphics{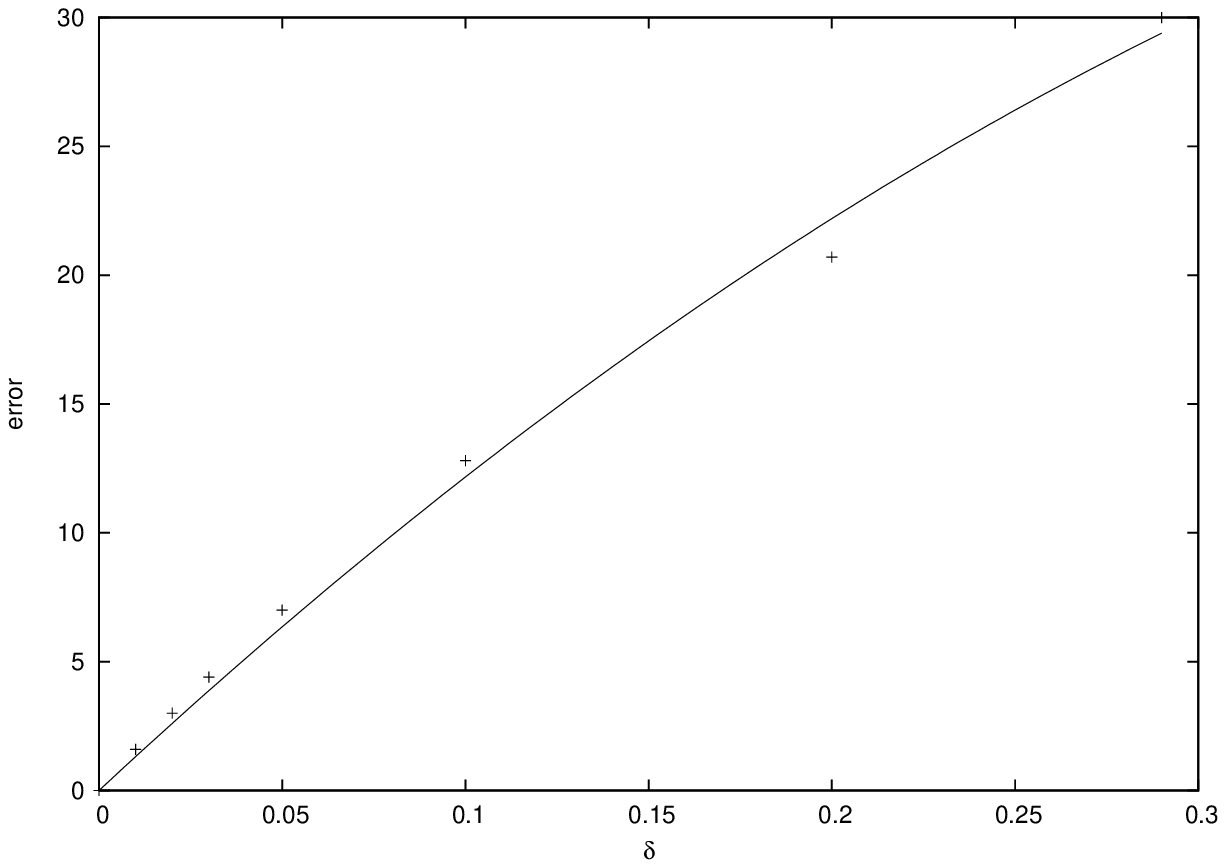}
\caption{}

\end{figure}
\newpage
\begin{figure}[ht]
\vskip 15truecm
 \includegraphics{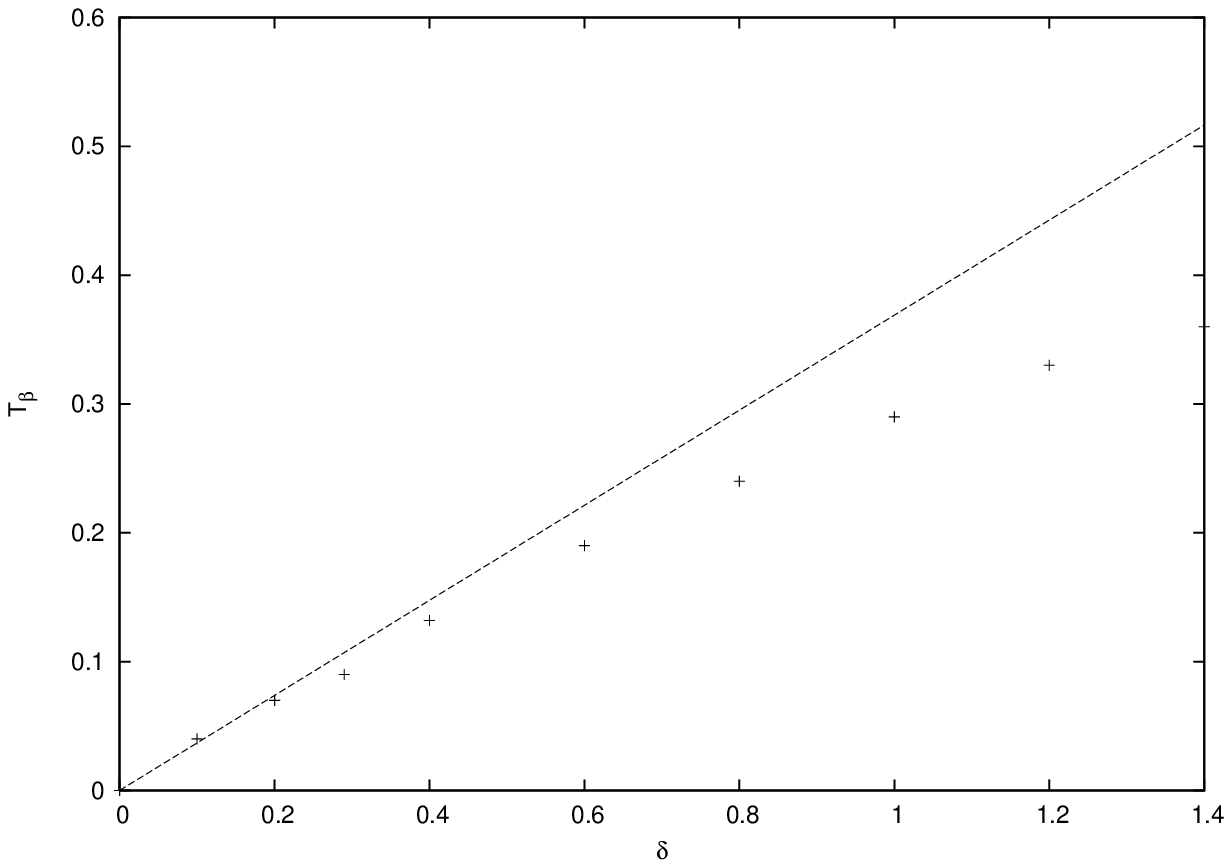}
\caption{}

\end{figure}
\newpage
\begin{figure}[ht]
\vskip 15truecm
 \includegraphics{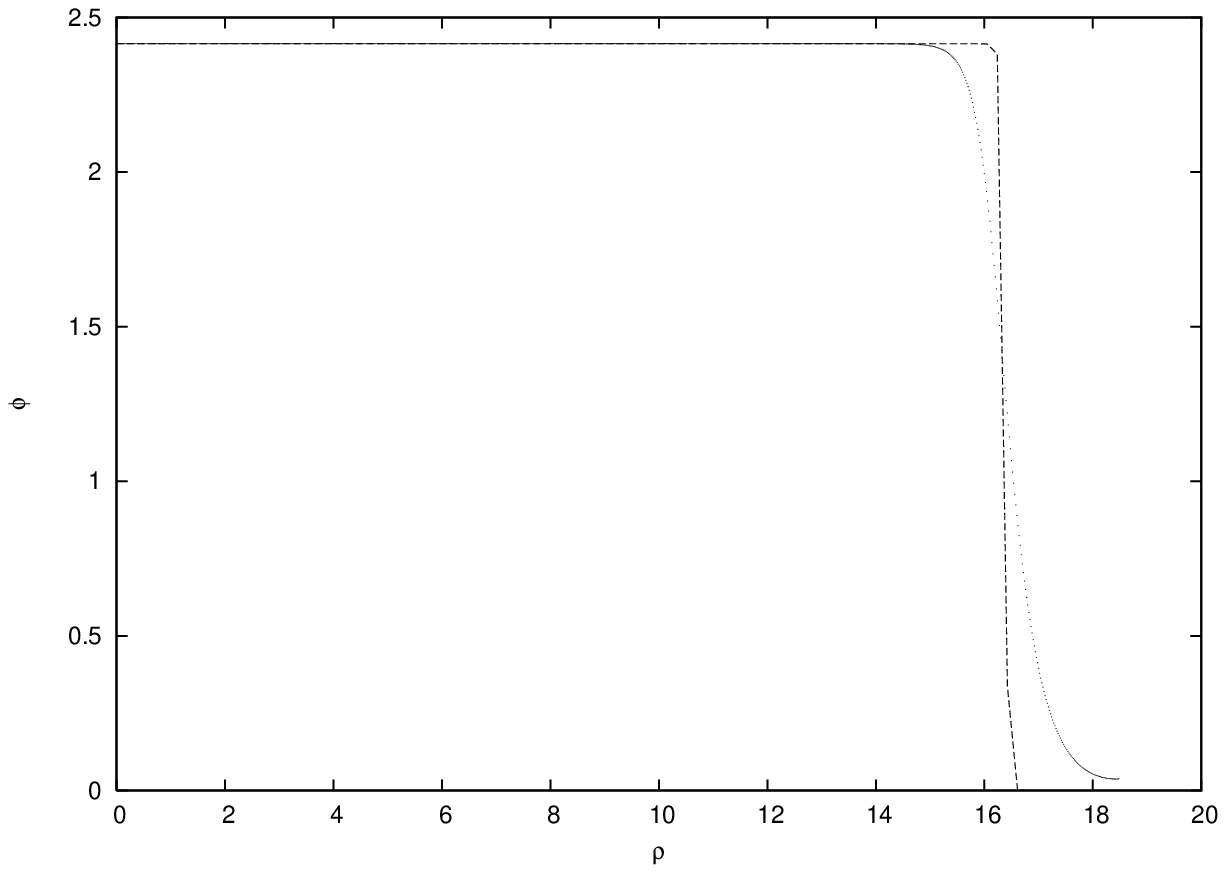}
\caption{}

\end{figure}
\newpage
\begin{figure}[ht]
\vskip 15truecm
 \includegraphics{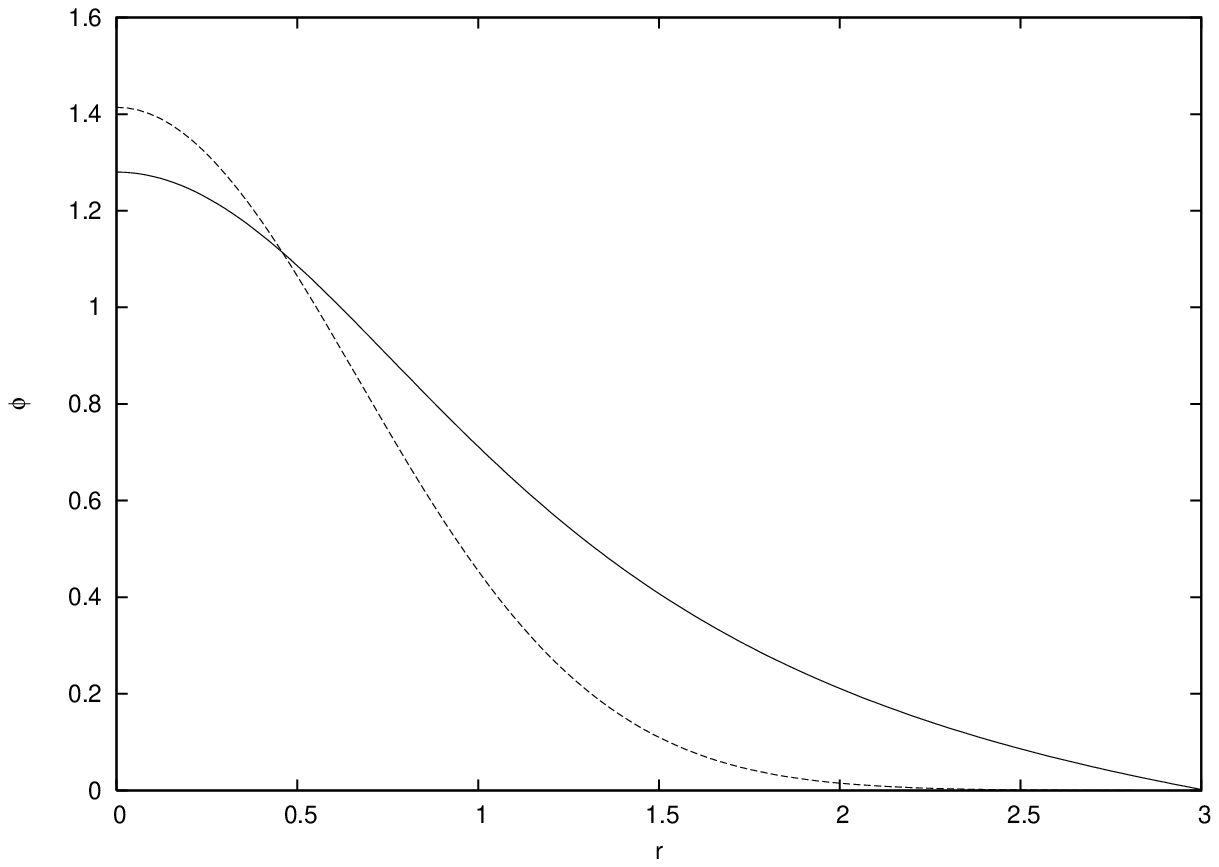}
\caption{}

\end{figure}

\end{document}